# Magnetic Fields in Earth-like Exoplanets and Implications for Habitability around M-dwarfs

Mercedes López-Morales[1,3], Natalia Gómez-Pérez[2,3], Thomas Ruedas[3]

[1]*Institut de Ciències de L'Espai (CSIC-IEEC), Barcelona, Spain*

[2]*Departamento de Física, Universidad de Los Andes, Bogotá, Colombia*

[3]*Carnegie Institution of Washington, Department of Terrestrial Magnetism, Washington D.C., USA*

e-mail: mlopez@ieec.uab.es; phone: +34 93 581 4369; fax: +34 93 581 4363

**Abstract** We present estimations of dipolar magnetic moments for terrestrial exoplanets using the Olson & Christiansen (2006) scaling law and assuming their interior structure is similar to Earth. We find that the dipolar moment of fast rotating planets (where the Coriolis force dominates convection in the core), may amount up to ~80 times the magnetic moment of Earth, $M_\oplus$, for at least part of the planets' lifetime. For very slow rotating planets (where the force of inertia dominates), the dipolar magnetic moment only reaches up to ~ 1.5 $M_\oplus$. Applying our calculations to currently confirmed rocky exoplanets, we find that CoRoT-7b, Kepler-10b and 55 Cnc e can sustain dynamos up to ~ 18, 15 and 13 $M_\oplus$, respectively. Our results also indicate that the magnetic moment of rocky exoplanets not only depends on their rotation rate, but also on their formation history, thermal state, age and composition, as well as the geometry of the field. These results apply to all rocky planets, but have important implications for the particular case of exoplanets in the Habitable Zone of M-dwarfs.

## Introduction

As progress is made towards identifying the first Earth-like planets, questions about their potential habitability have quickly emerged. Habitability critically depends on a number of factors that control the planet's surface conditions, among them the surface temperature and the shielding level against incident energetic charged particles.

The orbital star-to-planet distance defines the amount of irradiation that reaches the planets and also the magnitude of the tidal interaction effects which potentially result in synchronous rotation between the planet and the star, and therefore a slow-down of the planet's rotation rate. Those effects have been stated to affect habitability by inducing both uneven heating of the planet's atmosphere and surface, and a strong reduction of dynamo-generated magnetic shielding (e.g. Griessmeier et al. 2005; Griessmeier et al. 2009). Recent calculations by e.g. Correia et al. (2008), Barnes et al. (2009) and Heller et al. (2011) suggest that the planets may in fact end in stable non-synchronous rotation states if their orbits are slightly eccentric. In this work we explore the other effect, i.e. the potential generation of significantly strong dipolar magnetic moments even in slow-rotation conditions, as long as enough convection occurs in the planetary interiors.

**Magnetic Moment Generation Model**

Early planetary magnetic field theories focused on understanding magnetism in Solar System planets and satellites. Early model attempts (see Sano 1993 and references therein), concluded that the dipolar magnetic moment of planets, *M*, was proportional to some power law of their angular frequency, $\Omega$, or equivalently their rotational period, $P = 2\pi/\Omega$. More recently, convection has been found to also be an important parameter for dynamo scaling laws (Olson & Christensen 2006; hereafter OC06). Compared with previous models, OC06 mispredicts by less than 15% on average the observed magnetic moments of Solar System bodies, while all other models produce mismatches up to 268%. In addition, for example, all models except OC06 fail to predict the observed absence of magnetic field in Venus. Based on these comparisons, we assume that OC06 is the best currently available planetary magnetic moment scaling law, and adopt it for our calculations. The OC06 scaling law has an expression of the form

$$M \cong 4\pi r_o^3 \beta \left(\overline{\rho}_o / \mu_o\right)^{1/2} \left(F D\right)^{1/3},$$

where *M* is the magnetic dipole moment, $r_o$ is the planetary core radius, β is a fitting coefficient with a value of 0.1 – 0.2 deduced from numerical simulations, $\overline{\rho}_o$ is the bulk density of the fluid, $\mu_o$ is the magnetic permeability of the vacuum, *F* is the average convective buoyancy flux, and *D* is the thickness of the core rotating shell where convection occurs.

The most critical of those parameters is *F*, which is defined as the total convective heat flux, and represents the combination of both thermal and chemical convection in the planetary cores. *F* reflects the strength of the convection-driven dynamo, and can be expressed in terms of *D*, $\Omega$, and the local Rossby number, $Ro_l$, i.e. the ratio between the inertia and Coriolis forces, all normalized to the values of those parameters for Earth, as

$$\frac{F}{F_\oplus} = \left(\frac{Ro_l}{Ro_{l\oplus}}\right)^2 \left(\frac{D}{D_\oplus}\right)^{2/3} \left(\frac{\Omega}{\Omega_\oplus}\right)^{7/3}.$$

OC06 find that $Ro_l$ has to be ≤ 0.1 for the dynamo to generate a dipolar magnetic field. If $Ro_l > 0.1$, the resultant magnetic moment will be multipolar, with a weaker dipolar component of strength $M^- = 0.05 \cdot M$. Therefore, depending on the values of *D* and $Ro_l$, a given amount of convective heat flux will generate of dipolar or multipolar dynamo, depending on the planet's rotation rate $\Omega$. For our calculations we have assumed $D = r_o (1-0.35) \propto r_o$. The determination of *F* is difficult because it depends on planetary interior structure and composition properties, which are very uncertain. Therefore, for the calculations in this work we have assumed that $F \propto r_o^2$, i.e. the most efficient dynamo (Heimpel et al. 2005), for at least some period of the planet's lifetime, which allows us to estimate the maximum magnetic moment a planet can get depending on its rotation rate.

One of the other parameters defining interior structure that is needed for applying the OC06 equation is $r_o$. In the case of transiting planets it is possible to derive their average density from their absolute mass and radius measurements. However, their interior structure cannot be unequivocally defined with only that information. We have therefore assumed that the planets are stratified in two layers: a mantle and a core, and have an atmosphere and/or ocean/crust layer that amount to less than 1% of the total radius of the planets, i.e. within current observational errors. We use two different compositions for each layer; pure iron and an iron alloy with 20 mol percent iron sulfide (FeS) composition for the core, and pure olivine and perovskite+ferropericlase composition for the mantle. The pure olivine composition resembles the upper mantle of Earth, while the perovskite+ferropericlase composition resembles the lower mantle. We then use all possible combinations of those core and mantle compositions to integrate density and pressure along adiabatic profiles using the third-order Birch-Murnaghan equation of state to describe the pressure-density relation. Those integrations provide radius-density profiles for each planet, from which we derive the values of $r_o$, $\bar{\rho}_o$ and $D$ that are then used in the OC06 equation to estimate the planet's magnetic moment.

**Magnetic Moment Estimates for Exo-Earths**

Figure 1 shows results of our magnetic dipolar moment estimations for terrestrial planets with masses and radii up to 12 $M_{Earth}$ and 2.8 $R_{Earth}$, and with different rotation rates. We show the results of pure iron core, perovskite+ferropericlase mantle composition models, which give the lowest magnetic moment strength of all the core-mantle composition combinations we have studied, and therefore provide the most conservative dipolar magnetic moment estimates for our assumed planetary interior conditions. For reference we also include in the figure the location of the transiting low-mass planets GJ 1214b, CoRoT-7b, Kepler-10b and 55 Cnc e, and estimated locations of the non-transiting planets Gl 581 d and Gl 581g, this last one still unconfirmed. In the fast rotating regime, strong enough dipolar magnetic moments can be generated for all planets, and the surface of those planets could be well shielded if the interior conditions we are assuming are correct. In these cases, the dipolar magnetic moment can reach values up to 80 $M_\oplus$ for the most massive and densest planets. In the slow rotation regime, expected to be typical of planets in the classical Habitable Zone of M-dwarfs, we find that not all the planets will have dipolar magnetic moments strong enough to shield their surfaces. Planets with masses below ~ 2 $M_{Earth}$ will have $M < M_\oplus$ in all cases. However, planets with higher mass, and higher densities, can exhibit dipolar magnetic moments stronger than Earth's, even for very slow rotation rates.

**Conclusions**

The magnetic moment of a planet depends not only on its rotation rate $\Omega$, but also on its chemical composition and the efficiency of convection in its interior, given by $F$. If a planet is rotating 'fast enough', $\Omega$ establishes whether the dynamo is dipolar or multipolar, but the actual magnetic moment strength will also depend heavily on $F$. Our calculations suggest that rocky exoplanets might have strong magnetic fields to protect their surface against stellar and cosmic irradiation for at least part of their lifetimes, if convection is sufficiently efficient. Our models, however, do not account for changes of *F* and *D* with age, or for the effect of extreme external conditions, such as e.g. highly inhomogeneous heating or very strong stellar winds. These results are work in progress and a better understanding of the exoplanets' interior structure, tides, and energy transport mechanisms in rocky planets are necessary before further progress can be made.

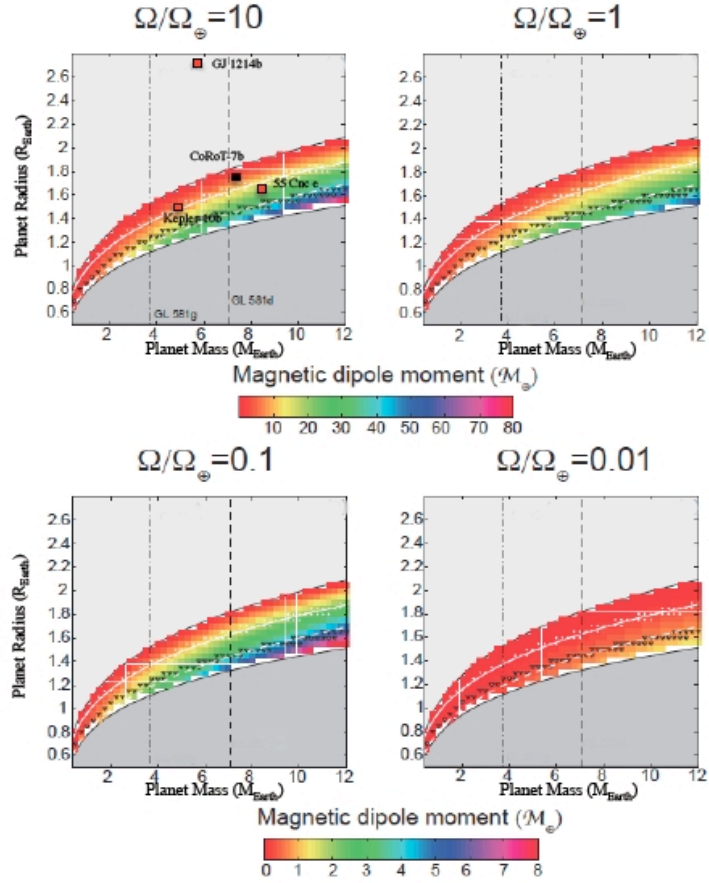

**Figure 1.** Predicted maximum dipolar magnetic moments as a function of planetary mass and radius, for four different rotation rates, assuming $F \propto r_o^2$ and perovskite+ferropericlase mantle and iron core compositions. The regions of the diagrams below the color points correspond to planets heavier than iron and are forbidden. The regions above the color points correspond to lower bulk density planets, i.e. not rocky, with substantial oceans (Neptune-like planets). For a more detailed description of these plots see *Gomez-Pérez, López-Morales & Ruedas 2011, submitted*.